\documentclass[onecolumn]{IEEEtran}
\usepackage{cite}
\usepackage{amsmath,amssymb,amsfonts}
\usepackage{algorithmic}
\usepackage{graphicx}
\usepackage{epstopdf}
\usepackage{textcomp}
\usepackage{multirow}
\usepackage{booktabs}
\usepackage{amsmath}
\usepackage{subfigure}
\usepackage{color}

\def\BibTeX{{\rm B\kern-.05em{\sc i\kern-.025em b}\kern-.08em
    T\kern-.1667em\lower.7ex\hbox{E}\kern-.125emX}}

\begin{document}
\title{Formulation of Single-Source Surface Integral Equation for Electromagnetic Analysis of Composite Penetrable Objects}
\author{Xiaochao Zhou, Zekun Zhu, and Shunchuan Yang
\thanks{This paper is a preprint of a paper submitted to IET Microwaves, Antennas \& Propagation. If
	accepted, the copy of record will be available at the IET Digital Library.}
\thanks{X. Zhou, Z. Zhu and S. Yang are with the School of Electronic and Information Engineering, Beihang University, Beijing, 100083, China (e-mail: zhouxiaochao@buaa.edu.cn, zekunzhu@buaa.edu.cn, scyang@buaa.edu.cn.)}
}

\maketitle

\begin{abstract}
This paper presents a new single-source surface integral equation (SS-SIE) to model composite penetrable objects. In the proposed formulation, the surface electric and magnetic fields on all interior boundaries are first eliminated through combining integral solutions inside each object. Then, by enforcing the surface electric fields in the original and equivalent configurations are equal to each other, an equivalent model with only the electric current density on the outermost boundaries is derived. Compared with other SIEs, like the PMCHWT formulation, all unknowns are residing on the outermost boundaries in the proposed formulation and therefore, less count of unknowns can be obtained. Finally, two numerical examples are carried out to validate the effectiveness of the proposed SS-SIE.
\end{abstract}

\begin{IEEEkeywords}
single-source, surface integral equation, equivalence theorem, differential surface admittance operator, composite

\end{IEEEkeywords}

\section{Introduction}
\label{sec:introduction}
Surface integral equations (SIEs) are widely used to model various electromagnetic problems, like scattering problems \cite{SCATTERING}, interconnect parameter extraction \cite{INTERCONNECT}. Since the unknowns only exist on the interfaces of different media, its overall count of unknowns is much smaller than that of the partial-differential-equation (PDE) based approaches, like the finite element method (FEM) \cite{FEM} and the finite-difference time-domain (FDTD) method \cite{FDTD}. 

The Poggio-Miller-Chang-Harrington-Wu-Tsai (PMCHWT) formulation \cite{PMCHWT} with both the electric and magnetic current densities is widely used to model composite structures. However, both the electric and magnetic current densities are required. Several single-source (SS) formulations are proposed to improve the efficiency \cite{SS1}\cite{SS2}. In \cite{DSAO}, a differential surface admittance operator (DSAO) is proposed to model interconnects and then, it has been extended to model dielectric objects \cite{DIELECTRIC} and composite structures \cite{APS_composite}. However, for objects embedded in multilayers, those formulations suffer from efficiency issues since the overall count of  unknowns increases rapidly with the count of interfaces. To solve this problem, \cite{ACES}-\cite{ZEKUN} have proposed an approach based on the DSAO to model objects embedded in multilayers, in which only a single electric current density is enforced on the outermost boundarys. Therefore, significant performance improvements in terms of the overall count of unknowns, memory consumption, and matrix conditioning can be obtained. However, the proposed approach does not take into account partially contacted composite penetrable objects. {Some modeling approaches for partially contacted objects have been proposed, like \cite{APS_composite} \cite{SS-SIE}. The SS-SIE proposed in \cite{APS_composite} applies the equivalence theorem to each object, which will lead to the existence of the electric current density inside and outside the interfaces of different objects. The macromodeling approach in \cite{SS-SIE} can significantly improve the computational efficiency of inhomogeneous antenna array. However, it requires both electric and magnetic current densities like the PMCHWT formulation.} In this paper, by carefully considering boundary conditions, an equivalent model with single electric current density enforced on the outermost boundaries is derived. The proposed single-source surface integral equation (SS-SIE) extends the capability of the approach in \cite{RESUBMIT} to solve the electromagnetic scattering problems by composite objects with partially connected boundaries. 

The paper is organized as follows. In Section II, the problem configurations and detailed formulations for partially contacted composite penetrable objects are presented. In Section III, the effectiveness of the proposed SS-SIE is investigated by two numerical examples. Finally, we draw some conclusions in Section IV.

\section{Methodology}
\subsection{Problem Configurations}
\begin{figure}
\centering
\subfigure[]{
\label{original_model} 
\includegraphics[width=0.3\textwidth]{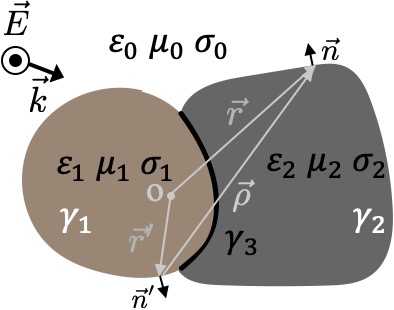}}
\hspace{0.1in}
\subfigure[]{
\label{equivalent_model} 
\includegraphics[width=0.34\textwidth]{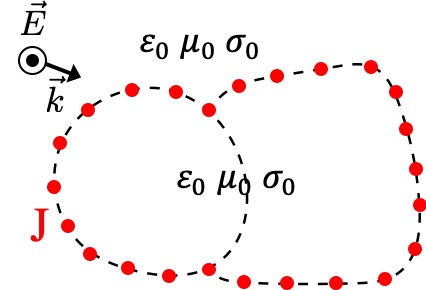}}
\caption{The problem configurations (a) the original composite object and (b) the equivalent model.}
\label{fig:subfig} 
\end{figure}

As shown in Fig. \ref{original_model}, a general composite object including two penetrable objects with different constant parameters is considered. In our proposed approach, an equivalent model with the electric current density is derived to make sure that the fields in the exterior region are exactly the same as those in the original model as shown in Fig. \ref{equivalent_model}. The composite structures interested in this paper are significantly different from those in \cite{RESUBMIT}\cite{ZEKUN}, where each penetrable object is fully embedded in another one and no partially contacted interfaces are available. Therefore, the proposed approach removes such constraint and makes the approach in \cite{RESUBMIT} suitable for any composite structures. 

To make our derivation concise, two penetrable partially touched objects are considered. The proposed approach can be easily extended to model other types of composite structures without any troubles. $\mu_1$, $\mu_2$, $\varepsilon_1$, $\varepsilon_2$, $\sigma_1$, and $\sigma_2$ denote the relative permeability, the relative permittivity, and the conductivity for the two penetrable objects and subscript 1, 2 are the object index, respectively. The boundary of object 1 is split into two parts, $\gamma_1$ and $\gamma_3$, and the boundary of object 2 is split into two parts, $\gamma_2$ and $\gamma_3$. $\gamma_3$ is shared by object 1 and object 2.

Our goal is to derive the equivalent model in Fig. \ref{equivalent_model}. In the equivalent model, the object is replaced by the background medium, and only the surface equivalent current density {\bf{J}} is enforced on the outermost boundaries $\gamma_1$ and $\gamma_2$ to keep the electromagnetic fields in the exterior region unchanged. According to the equivalence theorem, the surface equivalent electric current density ${\bf{J}} = {\bf{H}}_t - {\bf{\widehat H}}_t$ and the surface equivalent magnetic current density ${\bf{M}} = {\bf{E}}_t - {\bf{\widehat E}}_t$ should be introduced on the boundary of the equivalent structure. By make the tangential electric fields inside the boundary between the original and equivalent configurations equal to each other, namely ${\bf{E}}_t = {\bf{\widehat E}}_t$, the surface equivalent magnetic current density vanishes, and a SS-SIE can be obtained, as shown in \cite{DIELECTRIC}. Therefore, to derive the equivalent model in Fig. \ref{equivalent_model}, the relationship between the magnetic and electric fields on the boundary of the original and equivalent model are still required. In the next few subsections, we will derive the equivalent model in Fig. \ref{equivalent_model}.

\subsection{The Original Problem}
Without loss of generality, the composite structure does not include any sources. Therefore, the electric fields $E$ inside the structure in Fig. 1(a) must satisfy the homogeneous Helmholtz equation
\begin{equation}\label{Helmholtz}
\nabla^2 E + k_i^2 E = 0,
\end{equation}
where $k_i$ is the wavenumber inside object $i$.
According to the contour integral method \cite{INTEGRAL}, electric field $E_p$ on the boundary $\gamma_p$ ($p=1,2, 3$) can be expressed as 
\begin{equation}\label{SOLUTION OF HELMHOLTZ}
T E_p(\vec r) = \oint_{\gamma_p} \left[G_i(\vec r, \vec r') \frac {\partial E_p(\vec r)}{\partial n'}  - E_p(\vec r) \frac {\partial G_i(\vec r, \vec r')}{\partial n'} \right ]\,dr',
\end{equation}
where $G_i(\vec r, \vec r')$ is the Green function expressed as $G_i(\vec r, \vec r') = -jH_0^{(2)} (k_i \rho) / 4$, where $j = \sqrt {-1}$, $\rho = |\vec r - \vec r'|$ and $H_0^{(2)}$ is the \textit{zero}th-order Hankel function of the second kind, $T = 1/2$ when the source points $\vec r'$ and observation points $\vec r$ are located on the same boundary, otherwise, $T = 1$. The relationship between the electric and magnetic fields on $\gamma_1$ and $\gamma_3$ of object 1, and that on $\gamma_2$ and $\gamma_3$ of object 2 can be found through properly testing (\ref{SOLUTION OF HELMHOLTZ}). 

The electric fields $E_p$ and tangential magnetic fields $H_p$ on $\gamma_p$ are connected  by the Poincare-Steklov operator \cite{DSAO} as
\begin{equation}\label{STEKLOV}
H_p(\vec r) = \frac{1}{j\omega \mu_i}\frac{\partial E_p(\vec r)}{\partial n} \bigg |_{\vec r \in \gamma_p},
\end{equation}
where $\omega$ is the angular velocity.
$\gamma_1$, $\gamma_2$ and $\gamma_3$ are discretized into $N_1$, $N_2$ and $N_3$ line segments. The electric and magnetic fields are expanded by the pulse basis functions, and the coefficients are collected into the column vectors ${\bf E}_1$, ${\bf E}_2$, ${\bf E}_3$, ${\bf H}_1$, ${\bf H}_2$ and ${\bf H}_3$. By testing (\ref{SOLUTION OF HELMHOLTZ}) and (\ref{STEKLOV}) on $\gamma_1$ and $\gamma_3$ through the Galerkin scheme, we get
\begin{align}\label{E1}
{\bf L}_1{\bf E}_1 = {\bf P}_{11}{\bf H}_1+{\bf P}_{13}{\bf H}_3-{\bf U}_{11}{\bf E}_1-{\bf U}_{13}{\bf E}_3, \\
\label{E3}
{\bf L}_3{\bf E}_3 = {\bf P}_{31}{\bf H}_1+{\bf P}_{33}{\bf H}_3-{\bf U}_{31}{\bf E}_1-{\bf U}_{33}{\bf E}_3,
\end{align}
where matrix ${\bf L}_1$, ${\bf P}_{11}$, ${\bf P}_{13}$, ${\bf U}_{11}$, ${\bf U}_{13}$, ${\bf L}_3$, ${\bf P}_{31}$, ${\bf P}_{33}$, ${\bf U}_{31}$, and ${\bf U}_{33}$ are dimension of $N_1 \times N_1$, $N_1 \times N_1$, $N_1 \times N_3$, $N_1 \times N_1$, $N_1 \times N_3$, $N_3 \times N_3$, $N_3 \times N_1$, $N_3 \times N_3$, $N_3 \times N_1$, and $N_3 \times N_3$. Entries $(i,j)$ of matrix  ${\bf{L}}$, $\bf P$, $\bf U$ are given by 
\begin{align}\label{L}
 {\bf L}_{n_{(i,j)}} = \left \{
 \begin{aligned}
 &l_i &\quad (i = j) \\
 &0 &\quad (i \ne j)
\end{aligned},
\right.
\end{align}
\begin{align}
\label{P}
&{\bf P}_{{mn}_{(i,j)}} = \int_{{\gamma}_{m_i}} \int_{{\gamma}_{n_j}}\frac{\omega \mu_1}{2}H_0^{(2)}(k_1 \rho)\,dr'dr,\\
\label{U}
&{\bf U}_{{mn}_{(i,j)}} =  \frac{jk_1}{2} \int_{{\gamma}_{m_i}} \int_{{\gamma}_{n_j}}\frac{\vec \rho \cdot \hat n'}{\rho}H_1^{(2)}(k_1 \rho)\,dr'dr,
\end{align}
where $l_i$ is the length of the ${\it i}$-th segment on boundary $\gamma_n$, $\hat n'$ is the unit vector normal to the contour at the source point $\vec r'$, ${\vec \rho} = \vec r - \vec r'$, $\rho = |\vec \rho|$ and $H_1^{(2)}$ is the first order Hankel function of the second kind. By testing (\ref{SOLUTION OF HELMHOLTZ}) on $\gamma_2$ and $\gamma_3$, we can obtain the following two matrix equations for object 2
\begin{align}
\label{E3'}
{\bf L}_3{\bf E}'_3 = {\bf P}_{32}{\bf H}_2+{\bf P}'_{33}{\bf H}'_3-{\bf U}_{32}{\bf E}_2-{\bf U}'_{33}{\bf E}'_3,\\
\label{E2}
{\bf L}_2{\bf E}_2 = {\bf P}_{22}{\bf H}_2+{\bf P}_{23}{\bf H}'_3-{\bf U}_{22}{\bf E}_2-{\bf U}_{23}{\bf E}'_3,
\end{align}
where ${\bf E}_2$, ${\bf E}_3'$, ${\bf H}_2$ and ${\bf H}_3'$ are the expansion coefficient vectors of electric and magnetic fields on $\gamma_2$ and $\gamma_3$, and matrix ${\bf L}$, ${\bf P}$ and ${\bf U}$ also satisfy (\ref{L}), (\ref{P}) and (\ref{U}) with all constant parameters as $k_1$, $\mu_1$ replaced by those of object 2.

To eliminate additional unknowns on $\gamma_3$, the boundary conditions are given by
\begin{align}\label{BOUNDARY_E}
&{\bf E}_3 = {\bf E}'_3,\\
\label{BOUNDARY_H}
&{\bf H}_3 =- {\bf H}'_3.
\end{align}

By moving ${\bf E}_3$ and ${\bf E}_3’$ in (\ref{E3}) and (\ref{E3'}) to the left hand side (LHS) and using (\ref{BOUNDARY_E}), ${\bf E}_3$ and ${\bf E}_3’$ can be eliminated. Then, by further incorporating (\ref{BOUNDARY_H}), we get 
\begin{equation}\label{H3}
{\bf H}_3 ={{\bf C}_1} {\bf H}_1 + {{\bf C}_2} {\bf E}_1 + {{\bf C}_3} {\bf H}_2  + {{\bf C}_4} {\bf E}_2,
\end{equation}
where 
\begin{equation}
\begin{aligned}
&{{\bf C}_1} = {-{\bf B}_1^{-1}({\bf A}_1{\bf P}_{31})}, {{\bf C}_2} = {{\bf B}_1^{-1}({\bf A}_1{\bf U}_{31})}, \\
&{{\bf C}_3} = {{\bf B}_1^{-1}({\bf A}_2{\bf P}_{32})}, {{\bf C}_4}  = {-{\bf B}_1^{-1}({\bf A}_2{\bf U}_{32})}, \\
&{{\bf A}_1} = {({\bf L}_3 + {\bf U}_{33})^{-1}}, {{\bf A}_2} = {({\bf L}'_3 + {\bf U}'_{33})^{-1}}, \\
&{{\bf B}_1} = {{\bf A}_1{\bf P}_{33}+{\bf A}_2{\bf P}'_{33}}.
\end{aligned}
\end{equation}

By substituting (\ref{H3}) into (\ref{E3}), ${\bf E}_3$ can be written as
\begin{equation}\label{E3_END}
{\bf E}_3 = {{\bf D}_1}{\bf H}_1  + {{\bf D}_2}{\bf E}_1 + {{\bf D}_3}{\bf H}_2 + {{\bf D}_4}{\bf E}_2 ,
\end{equation}
where 
\begin{equation}
\begin{aligned}
& {{\bf D}_1} = {{\bf A}_1{\bf P}_{31} + {\bf A}_1 {\bf P }_{33} {\bf C}_1},\\
& {{\bf D}_2} = {-{\bf A}_1{\bf U}_{31} + {\bf A}_1 {\bf P }_{33} {\bf C}_2},\\
& {{\bf D}_3} = {{\bf A}_1 {\bf P }_{33} {\bf C}_3}, {{\bf D}_4} = {{\bf A}_1 {\bf P }_{33} {\bf C}_4}.
\end{aligned}
\end{equation}

By substituting (\ref{H3}), (\ref{E3_END}), (\ref{BOUNDARY_E}) and (\ref{BOUNDARY_H}) into (\ref{E1}) and (\ref{E2}), the relationship between the magnetic and electric fields on $\gamma_1$ and $\gamma_2$ can be expressed as
\begin{equation}\label{12}
\begin{aligned}
\underbrace{\left[ \begin{matrix}  {\bf M}_1 & {\bf M}_3  \\ {\bf F}_1 & {\bf F}_3  \\  \end{matrix} \right]}_{{\bf Q}_1} \left[ \begin{matrix} {\bf H}_1  \\  {\bf H}_2  \\ \end{matrix} \right]
 &=\underbrace{ \left[ \begin{matrix}  {\bf I} - {\bf M}_2 & - {\bf M}_4  \\ - {\bf F}_2 & {\bf I} - {\bf F}_4  \\  \end{matrix} \right]}_{{\bf Q}_2} \left[ \begin{matrix} {\bf E}_1  \\  {\bf E}_2  \\ \end{matrix} \right],
 \end{aligned}
 \end{equation}
where ${{\bf Q}_1}$ and  ${{\bf Q}_2}$ are square matrices with dimension of $(N_1 + N_2) \times (N_1 + N_2)$ and   $(N_1 + N_2) \times (N_1 + N_2)$, 
\begin{equation}
\begin{aligned}
&{{\bf M}_1} = {{\bf P}_{11} + {\bf P}_{13} {\bf C}_1 - {\bf U}_{13} {\bf D}_1}, \\
&{{\bf M}_2} = {- {\bf U}_{11} + {\bf P}_{13} {\bf C}_2 - {\bf U}_{13} {\bf D}_2},\\
&{{\bf M}_3} = {{\bf P}_{13} {\bf C}_3 - {\bf U}_{33} {\bf D}_3}, {{\bf M}_4} = {{\bf P}_{13} {\bf C}_4 - {\bf U}_{13} {\bf D}_4}, \\
&{{\bf F}_1} = {-{\bf {P}}_{23} {\bf C}_1 - {\bf U}_{23} {\bf D}_1 }, {{\bf F}_2} = {-{\bf {P}}_{23} {\bf C}_2 - {\bf U}_{23} {\bf D}_2 }, \\
&{{\bf F}_3} = {{\bf P}_{22} -  {\bf {P}}_{23} {\bf C}_3 - {\bf U}_{23} {\bf D}_3},\\ 
&{{\bf F}_4} = {- {\bf U}_{22} -  {\bf {P}}_{23} {\bf C}_4 - {\bf U}_{23} {\bf D}_4}.
 \end{aligned}
\end{equation}

By inverting the square matrix ${{\bf Q}_1}$, we obtain
\begin{equation}\label{12}
\begin{aligned}
\left[ \begin{matrix} {\bf H}_1  \\  {\bf H}_2  \\ \end{matrix} \right] &=\underbrace{ {\bf Q}_1^{-1} {\bf Q}_2}_{\bf Y} \left[ \begin{matrix} {\bf E}_1  \\  {\bf E}_2  \\ \end{matrix} \right],
\end{aligned}
\end{equation}
where $\bf{Y}$ is the surface admittance operator \cite{SKIN} to relate the electric and magnetic fields on $\gamma_1$ and $\gamma_2$ for the original object.

\subsection{The Equivalent Problem}
In the equivalent problem, the composite structure is replaced by its background medium and a surface equivalent current density {\bf{J}} is enforced on $\gamma_1$ and $\gamma_2$ to keep the fields in the exterior region unchanged in Fig. 1(b). The relationship between the electric and magnetic fields in the equivalent model can be obtained with similar manners in \cite{SKIN} and expressed as
\begin{equation}\label{13}
\left[ \begin{matrix} {\bf\widehat H}_1  \\  {\bf\widehat H}_2  \\ \end{matrix} \right] ={\bf\widehat Y} \left[ \begin{matrix} {\bf E}_1  \\  {\bf E}_2  \\ \end{matrix} \right],
\end{equation}
where ${\bf\widehat H}_1$, ${\bf\widehat H}_2$ are the coefficient vectors of magnetic fields on $\gamma_1$ and $\gamma_2$, respectively,  $\bf\widehat{Y}$ is the surface admittance operator for the equivalent model.

Therefore, according the surface equivalence theorem \cite{EQUIVALENCE}, the equivalent surface current density $\bf{J}$ on $\gamma_1$ and $\gamma_2$ can be expressed as
\begin{equation}\label{J}
\bf{J} = \left[ \begin{matrix} {\bf J}_1  \\  {\bf J}_2  \\ \end{matrix} \right] =\underbrace{ (\bf{Y} - \bf\widehat{Y})}_{{\bf Y}_s} \left[ \begin{matrix} {\bf E}_1  \\  {\bf E}_2  \\ \end{matrix} \right],
\end{equation}
where ${\bf J}_1$, ${\bf J}_2$ are the coefficient vectors of equivalent current densities on $\gamma_1$ and $\gamma_2$, respectively, and ${\bf Y}_s$ is the differential surface admittance operator.

\subsection{Scattering Modeling}

The scattering fields induced by ${\bf J}$ in the exterior region of the equivalent problem can be expressed as
\begin{equation}\label{E_SCATTERING}
{\bf{E}}^s(\vec r) = -j \omega \mu_0 \oint_{\gamma_1 + \gamma_2}{\bf{J}}(\vec r') G_0(\vec r, \vec r') \,dr'.
\end{equation}
Since the scattering electric field is in the background medium, the constant parameters should be $\varepsilon_0$, $\mu_0$ and $k_0$. The total fields expanded with the pulse basis functions in the exterior region can be expressed as the superposition of the scattering and incident fields
\begin{equation}\label{E_ALL}
{\bf{E}}(\vec r) = {\bf{E}}^s(\vec r) + {\bf{E}}^i(\vec r).
\end{equation}

Finally, through substituting (\ref{J}) and (\ref{E_SCATTERING}) into (\ref{E_ALL}), testing (\ref{E_ALL}) with the Galerkin scheme on boundary $\gamma_1$ and $\gamma_2$ and inverting the square coefficient matrix, we can obtain the electric field on $\gamma_1$ and $\gamma_2$  as
\begin{equation}\label{17}
{\bf E} = ({\bf I} - {\widehat {\bf P}}{\bf Y}_s)^{-1}{\bf E}^i,
\end{equation}
where $\bf{I}$ is an identity matrix with dimension of $(N_1+N_2) \times (N_1+N_2)$, and the element $(i,j)$ of $\widehat{\bf P}$ is expressed as ${\widehat {\bf P}}_{(i,j)} = \oint_{(\gamma_1+\gamma_2)_j} \frac{\omega_0 \mu_0}{2} H_0^{(2)}(k_0\rho)dr'$. Once the electric fields on the boundary $\gamma_1$, $\gamma_2$ are calculated, other interested parameters, like the radar cross section (RCS), near fields, are easy to be obtained.
\begin{figure}
	\centerline{\includegraphics[width=2.4in]{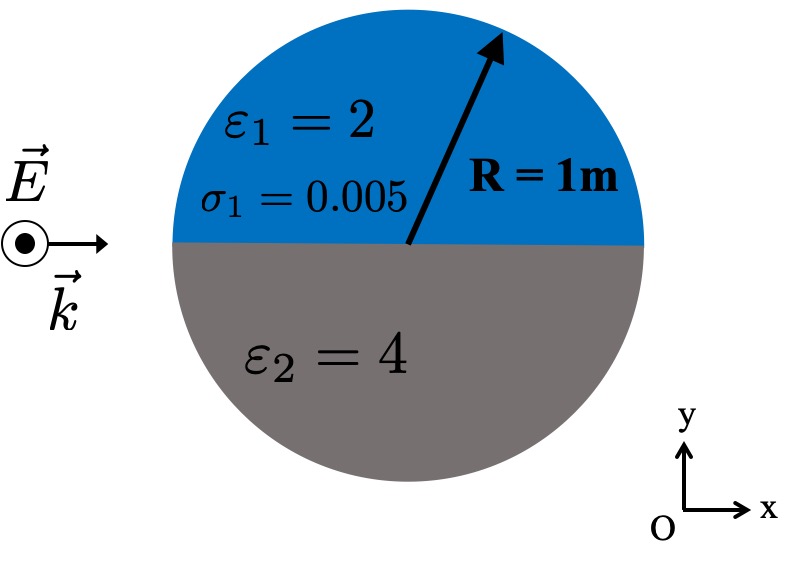}}
	
	\caption{A semi-contacted cylinder.}
	\label{example1_RCS_model}	
\end{figure}

\begin{figure}
	\centerline{\includegraphics[width=3.8in]{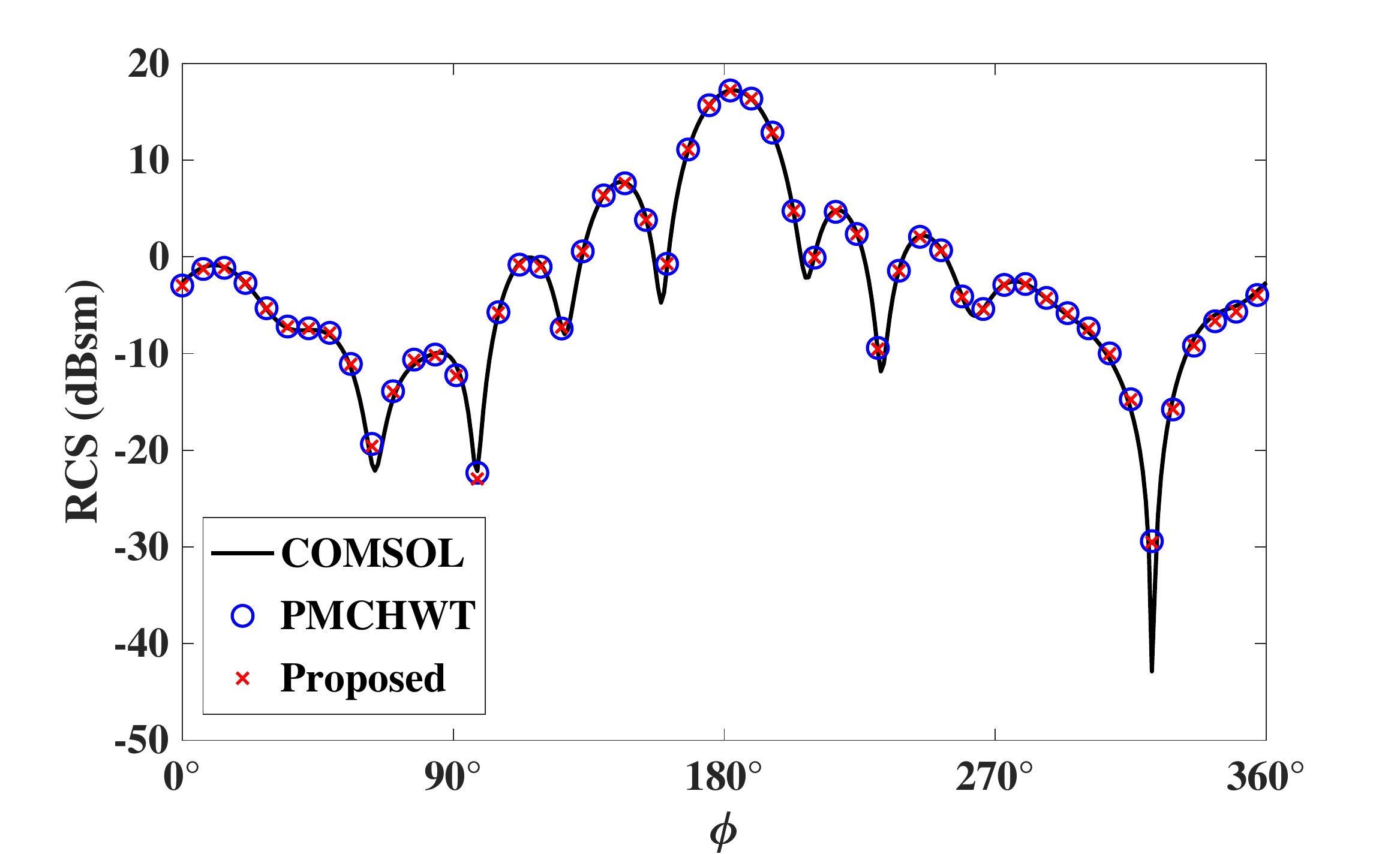}}
	
	\caption{RCS obtained from the COMSOL, the PMCHWT formulation, and the proposed approach.}
	\label{example1_RCS_model}	
\end{figure}

\section{Numerical Examples and Discussion}
\subsection{A Semi-contacted Cylinder}

\begin{figure}
	\centering
	\subfigure[]{
		\begin{minipage}[b]{0.32\textwidth}
			\includegraphics[width=1.2\textwidth]{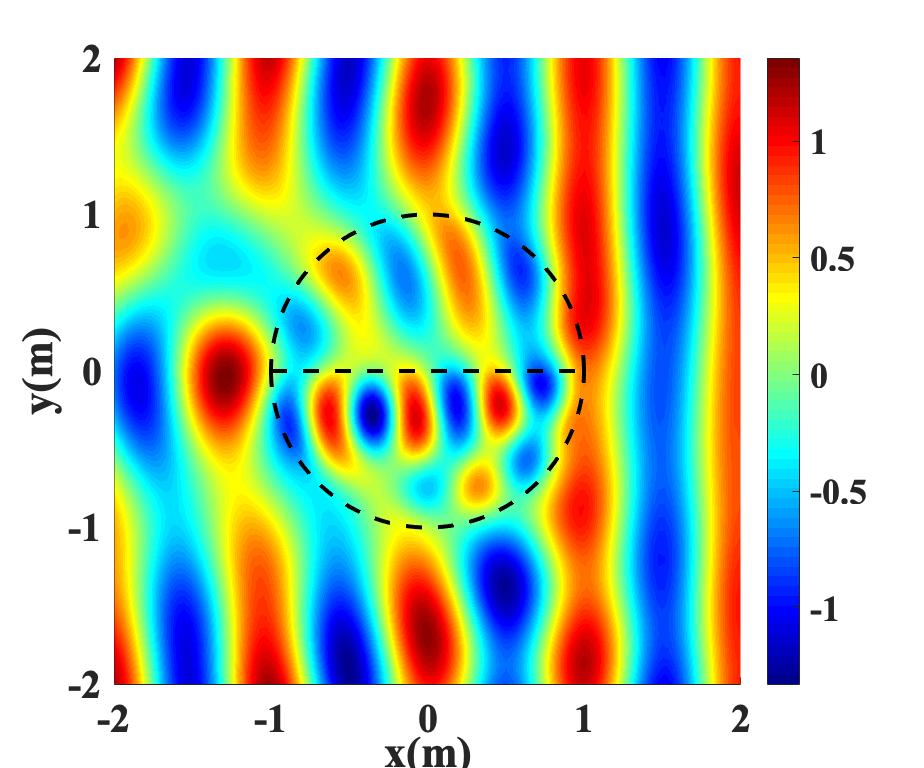}
		\end{minipage}
		\label{example1_electricalfield}
	}
	\hspace{1in}
	\subfigure[]{
		\begin{minipage}[b]{0.32\textwidth}
			\includegraphics[width=1.2\textwidth]{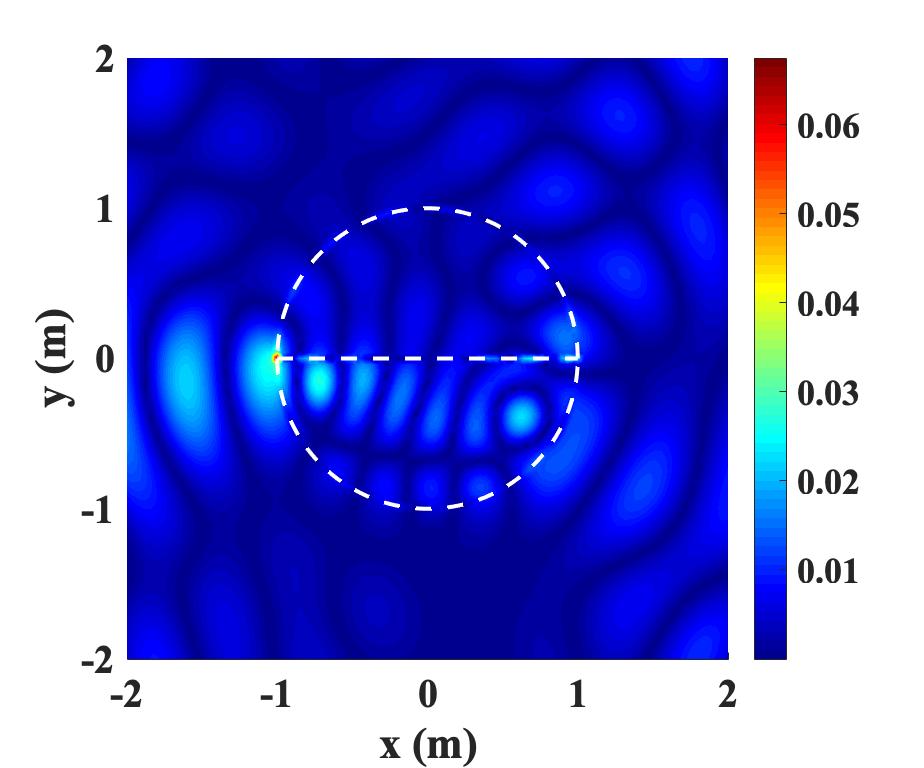}
		\end{minipage}
		\label{example1_electricalfield_error}
	}
	\caption{(a) The electric fields obtained from the proposed approach, and (b) the relative error of electric fields obtained from the COMSOL and the proposed approach in the near region of the composite object. The dotted line indicates the location of the original structure.}
	\label{example1}
\end{figure}

The first numerical example considered to verify the proposed approach is a composite structure including two semi-contacted cylindrical objects. The radius is $1$ m and the relative dielectric constant and the conductivity of the upper half-cylinder are 2 and 0.05, respectively. The relative dielectric constant of the lower half lossless cylinder is 4. The averaged segment length of $0.05$ m is used to discretize all the boundaries. A plane wave incidents along the positive $x$-axis with $f = 300$ MHz.

Fig. \ref{example1_RCS_model} shows the RCS obtained from the COMSOL, the PMCHWT formulation, and the proposed approach. The reference RCS is obtained from the COMSOL. It is easy to find that results obtained from the proposed approach show excellent agreement with those from the COMSOL and the PMCHWT formulation, which implies that the proposed approach can achieve accurate results as the traditional PMCHWT formulation. The PMCHWT formulation requires $332$ unknowns to solve this problem, while our proposed approach only requires $126$ unknowns with the same discretization, 38\% of unknowns in the PMCHWT formulation, which shows significantly less count of unknowns.

Fig. \ref{example1_electricalfield} illustrates electric fields in the near region of the composite object obtained from the proposed approach. Fig. \ref{example1_electricalfield_error} presents the relative error of electric fields. The relative error is calculated by $\left |  {\text{E}^\text{cal}-\text{E}^\text{ref}} \right | / \left | {\text{E}^\text{ref}} \right |$, where $\text{E}^\text{ref}$ and $\text{E}^\text{cal}$ denote the reference fields obtained from the COMSOL and the calculated fields from the proposed approach, respectively. As shown in Fig. \ref{example1_electricalfield_error}, the relative error in most regions is less than 3\%. Only a few points show slightly large relative errors. Therefore, numerical results show that the proposed approach can accurately calculate near fields.

\begin{table}[]
\centering
\caption{Comparison of the condition number between the proposed approach and the PMCHWT formulation}
\begin{tabular}{l|c|c}
\toprule
\midrule
\multicolumn{2}{c|}{Mertic}                     &  Condition Number \\ \hline
\multirow{6}{*}{{Proposed}} & ${{\bf A}_1}$                   & 1     \\ 
                          & ${{\bf A}_2}$                    & 1     \\  
                          & ${{\bf B}_1}$                    & 11.48 \\ 
                          & ${{\bf Q}_1}$                   & 19.02 \\ 
                          & ${{\bf P}_{out}}$               & 33.31 \\  
                          & ${\bf I} - {\widehat {\bf P}}{\bf Y}_s$(final matrix) & {\color{red} 70.36} \\ \hline
\multicolumn{2}{c|}{ PMCHWT formulation}                     & {\color{blue} 4,427}  \\ \hline
\bottomrule
\end{tabular}
\end{table}

In Table I, we list the 2-norm condition numbers of various matrices obtained through the Matlab command \text{``cond(`matrix',2)''}, which are required to get the inverses in the proposed approach and the PMCHWT formulation. Table I clearly shows that the maximum condition number in our method is 70.36, while the condition number in PMCHWT is 4,427. It can be found that the proposed SS-SIE can significantly improve the conditioning of the final linear system. In addition, it is easy to find that the matrix condition numbers of all the intermediate matrices which are required to be inverted are quite small, which implies that quite good convergence can be obtained if iterative algorithms are used to invert those matrices.

\subsection{A Composite Cuboid}
\begin{figure}	
 	\centerline{\includegraphics[width=3.3in]{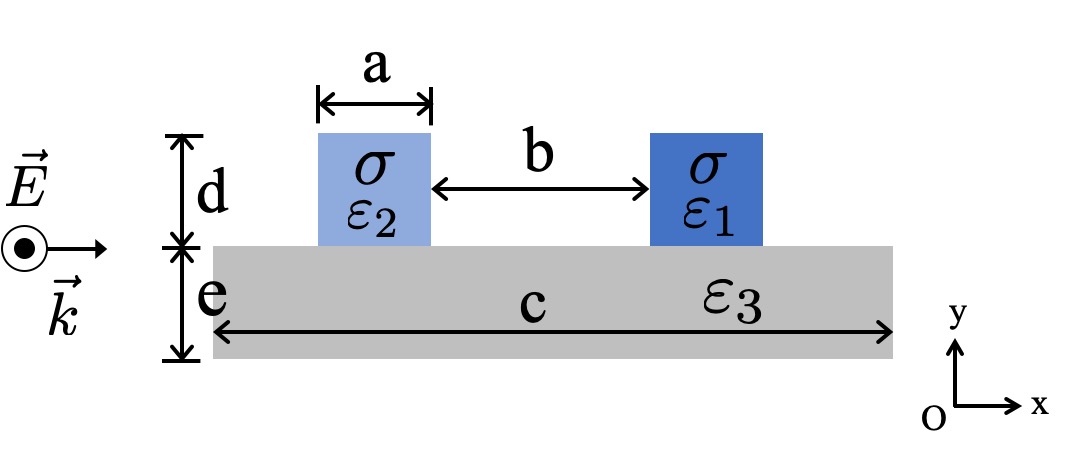}}
 	
 	\caption{A composite cuboid.}
 	\label{example2_model}
 \end{figure}

A composite penetrable object including three dielectric cuboids as shown in Fig. \ref{example2_model}. The two small cuboids are symmetrically placed with constant parameters $\varepsilon_{r_1}=2$, $\varepsilon_{r_2}=3$, $\sigma = 0.005$, $a=d=0.5$ m and $b=1$ m. The width and length of the dielectric cuboid object below are $e=0.5$ m and $c=3$ m, and the constant parameter $\varepsilon_{r_3}=5$. A plane wave incidents along the positive $x$-axis with $f = 300$ MHz.
 
 \begin{figure}	
 	\centerline{\includegraphics[width=3.8in]{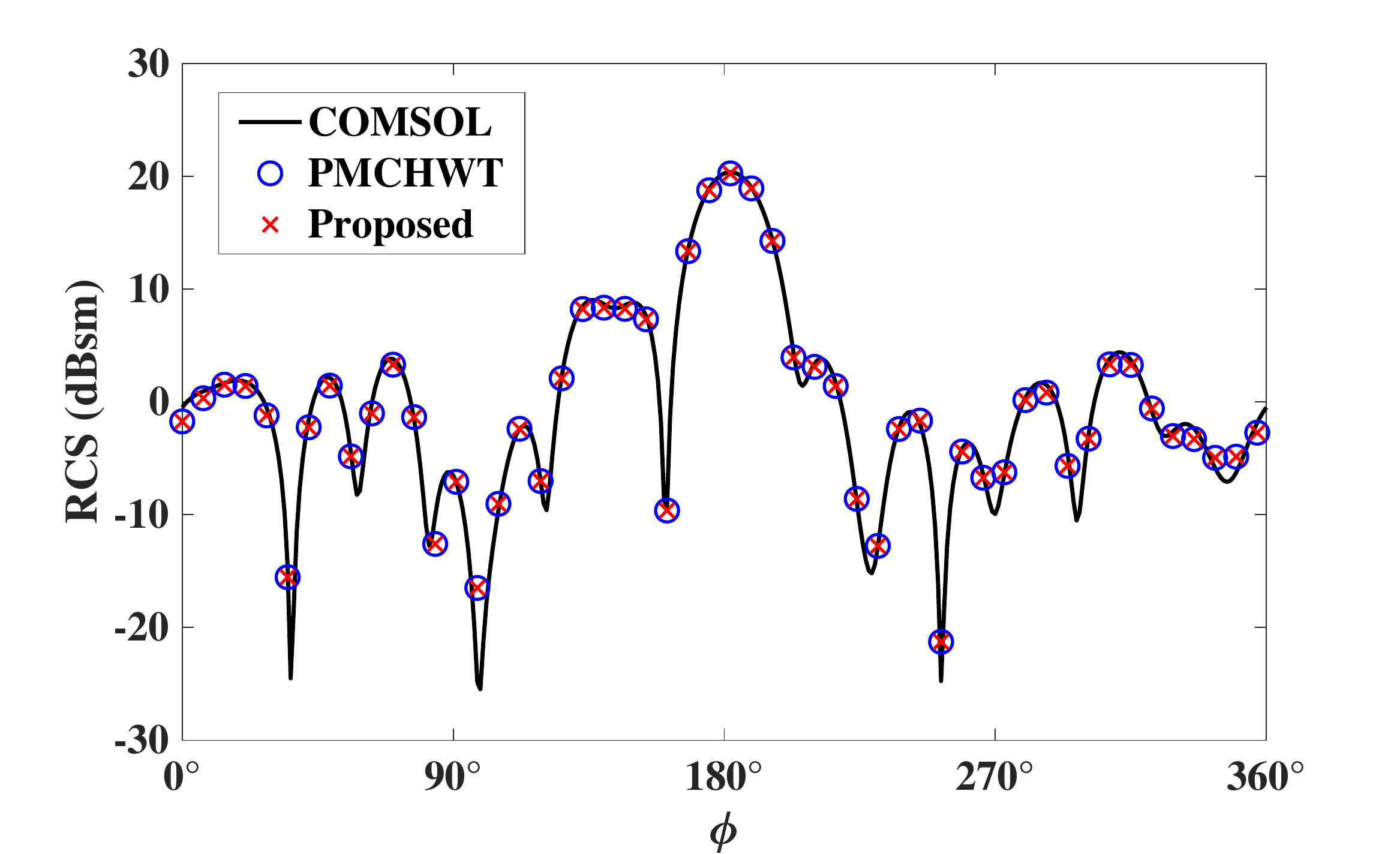}}
 	
 	\caption{RCS obtained from the COMSOL, the PMCHWT formulation, and the proposed approach.}
 	\label{example2_RCS_model}
 \end{figure}

 \begin{figure}	
 	\centerline{\includegraphics[width=3.6in]{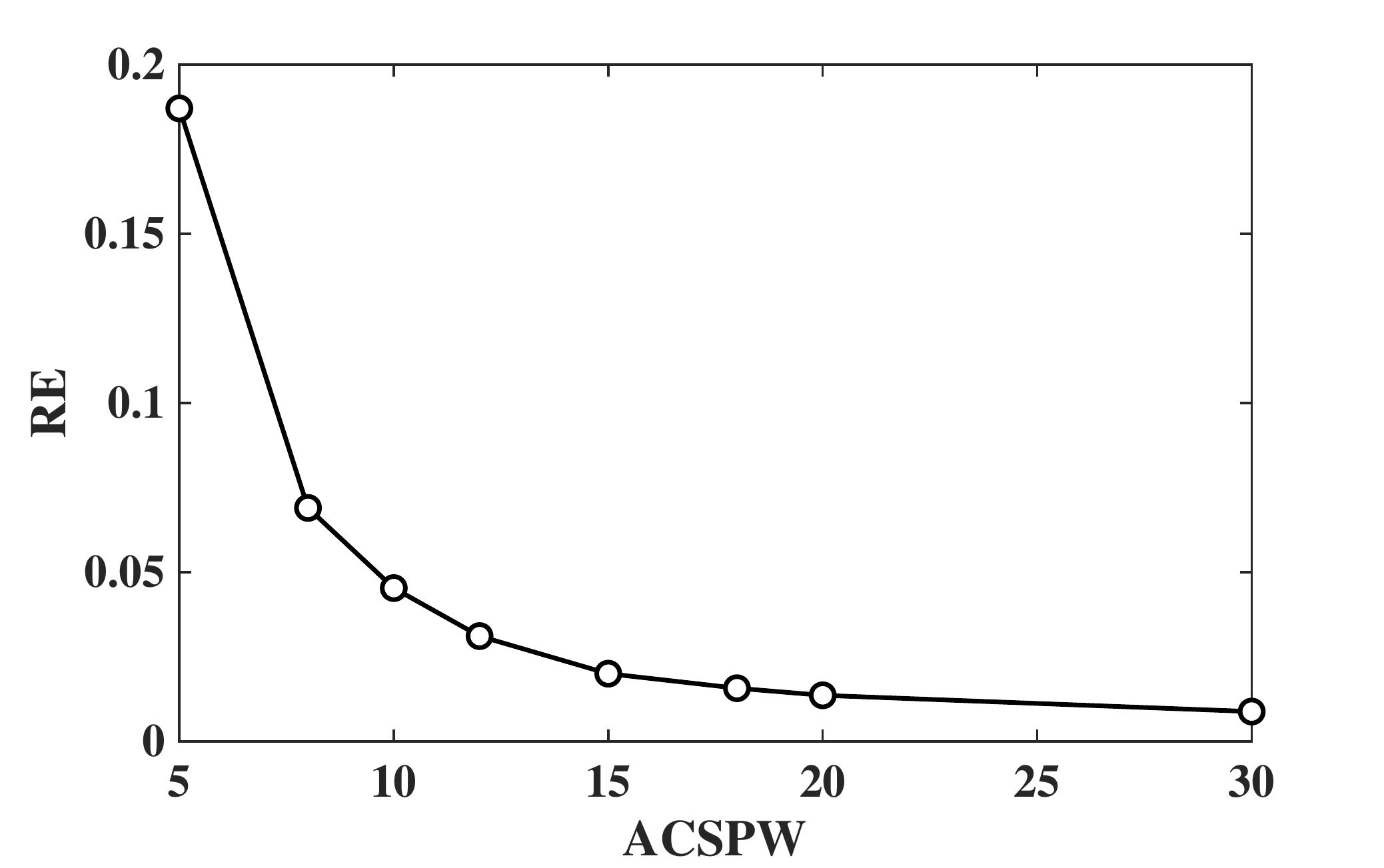}}
 	
 	\caption{The RE of the RCS verse ACSPW.}
 	\label{example2_ACSPW}
 \end{figure}
Fig. \ref{example2_RCS_model} shows the RCS from the COMSOL, the PMCHWT formulation, and the proposed approach. Results obtained from the proposed approach show excellent agreement with those from the COMSOL, the PMCHWT formulation. It implies that the proposed approach can obtain accurate results with non-smooth surfaces. The PMCHWT formulation requires $520$ unknowns to solve this problem. However, our proposed approach only requires $240$ under the same boundary partition, 46\% of unknowns in the PMCHWT formulation. Fig. \ref{example2_ACSPW} shows the relative error (RE) of RCS verse the average count of segments per wavelength (ACSPW) in the free space. The relative error is defined as 
\begin{equation}
\text{RE} = \frac{\sum_{i} \left \| \text{RCS}^{\text{cal}}_i - \text{RCS}^{\text{ref}}_i \right \| ^ {2}}{ \sum_i \left \| \text{RCS}^{\text{ref}}_i \right \| ^2 },
\end{equation}
where $\text{RCS}^{\text{cal}}$ is the calculated RCS from the proposed approach and $\text{RCS}^{\text{ref}}$ is obtained from the COMSOL with fine enough mesh. As shown in Fig. 7, when ACSPW is less than 15, the RE fast decreases as ACSPW increases. When ACSPW further increases, the RE slowly decreases. This is expected since only low order pulse basis functions are used in our implementations.

\begin{figure}	
	\centerline{\includegraphics[width=3.8in]{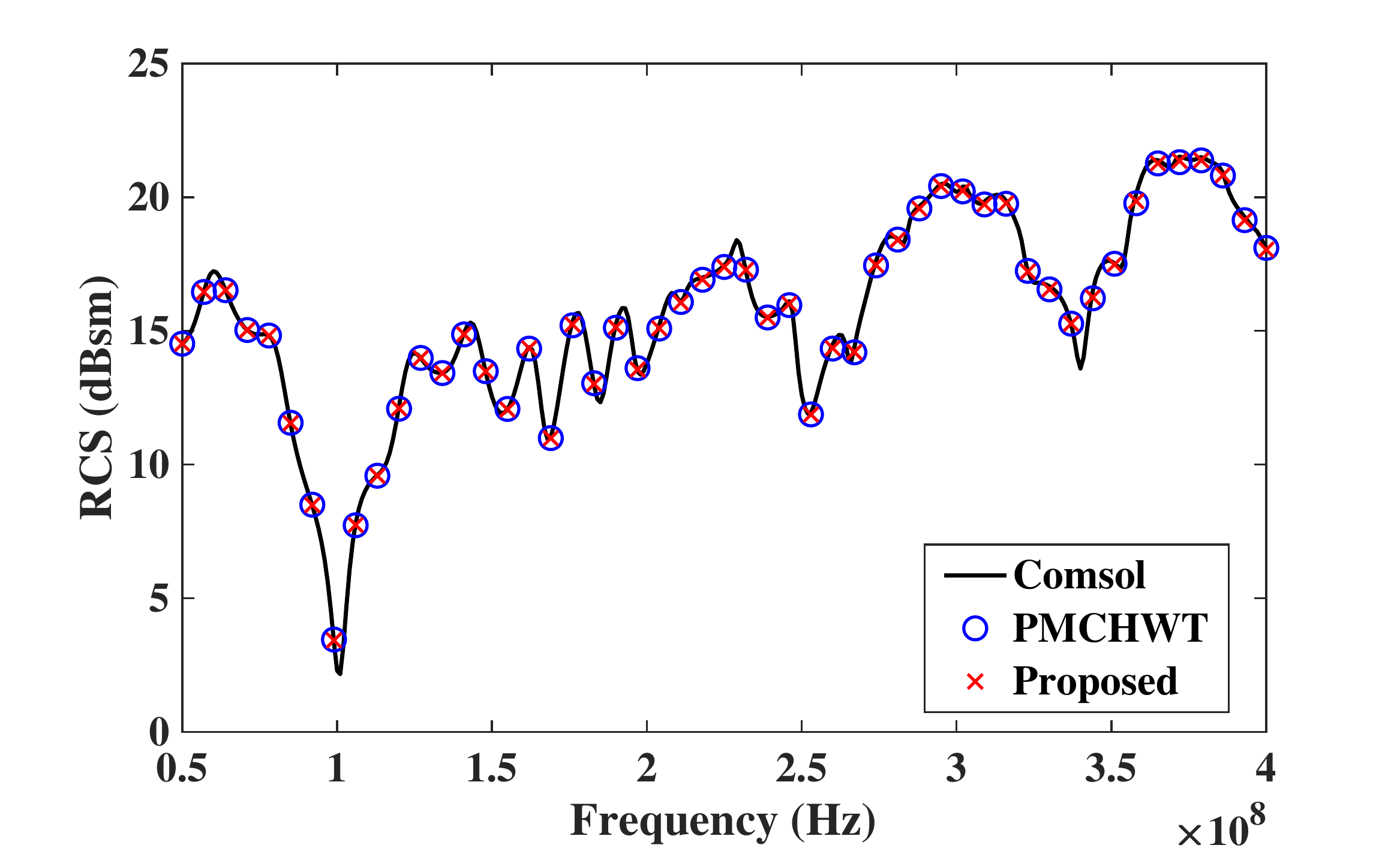}}
	
	\caption{The monostatic RCS from the COMSOL, the PMCHWT formulation and the proposed approach.}
	\label{example2_frequency}
\end{figure}

Fig. 8 shows the monostatic RCS obtained from the three approaches from 50 MHz to 400 MHz. It is found that results obtained from the proposed approach show excellent agreement with those from the PMCHWT formulation and the COMSOL in the wideband frequency range. 

\begin{table}[]
\centering
\caption{Comparison of the condition number between the proposed approach and the PMCHWT formulation}
\begin{tabular}{l|c|c}
\toprule
\midrule
\multicolumn{2}{c|}{Mertic}                     &  Condition Number \\ \hline
\multirow{11}{*}{{Proposed}} & ${{\bf A}_1}$                   & 1     \\ 
                          & ${{\bf A}_2}$                    & 1     \\  
                          & ${{\bf A}_3}$                    & 1     \\ 
                          & ${{\bf A}_4}$                    & 1     \\ 
                          & ${{\bf B}_1}$                    & 7.86 \\ 
                          & ${{\bf B}_2}$                    & 7.86 \\ 
                          & ${{\bf N}_1}$                    & 1 \\ 
                          & ${{\bf N}_2}$                    & 1\\ 
                          & ${{\bf S}_1}$                   & 475.87 \\ 
                          & ${{\bf S}_{out}}$               & 59.72 \\  
                          & ${\bf I} - {\widehat {\bf P}}{\bf Y}_s$(final matrix) & {\color{red} 5,001} \\ \hline
\multicolumn{2}{l|}{ PMCHWT formulation}                     & {\color{blue} 13,627}  \\ \hline
\bottomrule
\end{tabular}
\end{table}

In Table II, we listed the 2-norm condition numbers  of matrices, which are required to get the inverses in the proposed approach and the PMCHWT formulation. It should be noted here that the matrices in the table are not given specific expressions in the paper, which need to be derived with the same derivation steps as above. Table II clearly shows that the maximum condition number in our approach is 5,001, while the condition number in PMCHWT is 13,627. Similar to the previous numerical example, the proposed approach shows significant improvement in terms of the condition number.

\section{Conclusion}
In this paper, we proposed a SS-SIE for partially contacted composite penetrable objects. In the proposed approach, the composite object is replaced by the background medium and only the electric current density incorporated with the DSAO is enforced on the outermost boundaries. Compared with the traditional PMCHWT formulation, less than half of the number of unknowns is required in the proposed approach. In addition, it extends the capability of the approach in \cite{RESUBMIT} to handle partially contacted composite objects and then make it to a general approach which can deal with any type of composite structures. Numerical examples finally validate the accuracy and numerical convergence of the proposed formulation.

\end{document}